\documentclass[prd,twocolumn,preprintnumbers]{revtex4}
\usepackage{amsmath}
\usepackage{mathrsfs} 
\usepackage{epsfig}

\newcommand{\scri}{\mathscr{I}}
\newcommand{\secref}[1]{Sec.~\ref{sec:#1}}
\def\half{\tfrac{1}{2}}           

\begin{document}

\preprint{gr-qc/0603034}
\preprint{UM-PP-06-nn}
\preprint{AEI-2006-062}

\begin{abstract}
We study the numerical propagation of waves through future null infinity
in a conformally compactified spacetime.  We introduce an artificial
cosmological constant, which allows us some control over the causal
structure near null infinity.  We exploit this freedom to ensure that all light
cones are tilted outward in a region near null infinity, which allows us
to impose excision-style boundary conditions in our finite difference code.
In this preliminary study we consider electromagnetic waves propagating in
a static, conformally compactified spacetime.
\end{abstract}

\date{16 July 2006}                                                

\title{Excising das All: Evolving Maxwell waves beyond scri}
\author{Charles W. Misner}
\affiliation{Department of Physics, University of Maryland, College Park,
Maryland 20742-4111}
\affiliation{Albert-Einstein-Institut, Max-Planck-Institut
f\"ur Gravitationsphysik, Am M\"uhlenberg 1, D-14476 Potsdam, Germany
}
\author{James R. \surname{van Meter}}
\affiliation{Laboratory for Gravitational Astrophysics,
NASA Goddard Space Flight Center, Greenbelt, Maryland 20771}
\author{David R. Fiske}
\affiliation{Intelligent Systems Division, Decisive Analytics Corporation,
1235 South Clark Street, Arlington, Virigina 22202}
\affiliation{Laboratory for Gravitational Astrophysics,
NASA Goddard Space Flight Center, Greenbelt, Maryland 20771}
\maketitle

\section{Introduction}

The motivation for the present work is the matching of a computational model of a
source of gravitational waves (such as a compact object binary) to the
cosmological environment through which any emitted signal travels to a
detector.
The transmission phase is well understood and is essentially the
propagation of a linearized gravitational wave through a background
cosmology which includes such effects as a cosmological redshift and
gravitational lensing due to intervening masses, plus the kinetic
effects of the detector's motions.
For these purposes the geometrical optics approximation is often
adequate.
Although implementing this transmission model in observatory data
analysis demands skill, it is simpler than the electromagnetic case
where interactions with dust, gas and plasmas en route must also be
considered.
In this large scale background the source can be taken to be a point
source of linearized gravitational waves having some specific antenna
patterns and wave forms.
The more unsettled question then is how to extract from a detailed model
of a wave generation system the appropriate linearized wave.
The most common assumption is that the wave generator can be modeled
as existing in an asymptotically flat spacetime where waves propagating
out to future null infinity (scri+ or $\scri^+$) will be described
there by linearized gravitation, originating from a point source
in the cosmological context.
This is clearly justified by the huge difference in scales between
the size of the wave generator and the radius of curvature of its
surrounding spacetime (the cosmos or a galaxy or stellar cluster).
The question then is how to efficiently calculate and extract the
waveforms at $\scri^+$.

In the usual Cauchy evolutions on asymptotically flat spacelike
hypersurfaces, the waves must be recognized at large distances from the
source, possibly a long time after the source motions have ceased
violent activity.
One way to overcome this time lag is through the use of a retarded time
coordinate. This works well in spherical symmetry \cite{HM:retd,
BST:retd}, but is poorly defined otherwise.
The Pittsburgh group has used such characteristic evolution at larger
distances while matching to more common Cauchy evolution in the interior
\cite{winicour:ccm}.
The use of hyperboloidal time slices (some generalization of
hyperboloids in Minkowski spacetime) appears attractive as a smooth
transition from conventional time slices in central regions to
asympotically null spacelike hypersurfaces toward infinity.
Were such a slicing to be combined with a straightforward Cauchy
evolution, the Courant time step condition would defeat the advantages
of approximating retarded time, as the nearly infinite coordinate speed of light
toward infinity would require infinitesimal time steps (since the Courant
condition generally requires that $\Delta t/\Delta x < 1/c$ for numerical stability).
Thus hyperboloidal slicings effectively require a conformal
compactification which makes the coordinate speed of light finite again
\cite{MSL:KITP}.

In addition to the use of hyperboloidal time slices and conformal
compactification, we suggest two additional tools to ameliorate the
extraction of outgoing waves in numerical evolutions.
One is the use of artificial cosmology to convert the null hypersurface
$\scri^+$ into a spacelike hypersurface in the future of the initial
Cauchy hypersurface as proposed in \cite{misner:deserFest}.
This procedure fattens $\scri^+$ into a shell of spacelike hypersurfaces
by changing the value of the cosmological constant, plausible for the
present universe, by many orders of magnitude so that the de~Sitter
horizon is located, not at cosmological distances, but merely well
beyond the region where the wave generating mechanism can be influenced
by it.
In this sense it is similar to the computational tool of artificial
viscosity which fattens a shock from its physical thickness on the
order of the molecular mean free path into a much larger length which is
still small in the range of scales that otherwise are met in the
physical problem at hand.
The second additional tool is the tilting of the null cones governing
wave propagation in ways that ease the computational behavior in the
unphysical (`das All') region beyond the de~Sitter horizon and beyond
$\scri^+$.

The German noun ``das All'' refers to the Universe, but sometimes in a sense  
unfamiliar in English where it translates better as ``outer space.''  A
spacecraft, after launch, might be described as \emph{entering} das All.   
In the computations reported here, it is precisely this outer space --- the
physical details far from a source of gravitational (or in our case
electromagnetic) waves --- which is excised from the computational model
in analogy to the way the internal mysteries/singularities of a black hole
are often excised in computational models.  We develop and implement the 
proposals made earlier in 
Refs.~\cite{MSL:KITP,misner:deserFest} and suggested above to
obtain wave forms promptly at infinity by using hyperboloidal time slices
in a Cauchy evolution,                                                       
to avoid infinitesimal time steps through conformal compactification,        
to convert outer boundary conditions to excision                              
boundary conditions by using an artificial cosmological constant, and
to defend the physical domain against computational boundary errors by
tilting the light cones further outward in the region beyond the outer
horizon.
The aim of this paper is to explore the application of this set of           
boundary treatment tools in a low cost application in order to see           
whether there are unexpected impediments to its application, and to get      
guidance for the best approaches for its application to larger problems.     

The use of hyperboloidal slicings in numerical relativity has been
carefully studied (e.g.\ Refs.~\cite{sascha:tuebingen,hf:sh,winicour:ccm,Frauendiner:2004lrr}),  
and there are existing numerical results generated on compactified
domains (e.g.\ Ref.~\cite{Pretorius:2005gq}).  This paper contributes to this
literature by combining these two techniques in a spacetime with an
artificial cosmological structure designed to make it easier to construct
boundary conditions at $\scri^+$, and by studying the effectiveness of
modifying the light cone structure in the region of the analytically
continued spacetime exterior to $\scri^+$ to prevent non-physical, incoming
radiation from piling up on the non-physical side of $\scri^+$.
That said, our approach to the problem does not require special  
modifications to the form of the differential equations solved to model
the physical problem.  It should be compatible with
existing numerical codes desgined to solve the initial problem, 
described by some variant of the the Einstein equations in 
an Arnowitt-Deser-Misner 3+1 decomposition \cite{adm}.

Our work is also complementary to community wide efforts to use mesh
refinement techniques in numerical relativity.  Several groups
have recently and successfully incorporated mesh refinement techniques into
their Einstein solvers (i.e.\ Refs.\
\cite{brug:fmrexc,goddard:waves,goddard:puncture,goddard:wavezone,
carpet:intro,maya:waves,Pretorius:2005gq,brug:orbits,
goddard:extraction,goddard:merger}).  This technology allows groups 
to push their computational boundaries to very large coordinate distances,
reducing the effects of incoming radiation and constraint violating modes
introduced by applying boundary conditions at a finite distance.  

While mesh refinement has greatly increased the power of modern codes, it also has
limitations. 
It is now fairly easy to surround a computational domain with coarser
and coarser grids, but at some point the coarse grids cannot resolve a
wavelength, see, e.g.\ Refs.~\cite{goddard:wavezone} and \cite[Section
4.3]{fiske:phd}. 
They can, however, insulate the active, wave generation regions of the
computational domain from boundary problems for a limited period of
time. 
During this period waveforms have been successfully extracted from inner
regions as small as 2.4 wavelengths from the center
\cite{goddard:merger}.
The hyperboloidal slicings considered here and elsewhere have the
advantage that they asymptote to the outgoing light cones in the
spacetime. 
Outgoing waves, therefore, appear asymptotically constant and require
very few points to resolve them all the way to $\scri^+$.

The rest of the paper is organized as follows: In \secref{formalism}, we
briefly summarize the formalism we use in our code, both for the
background spacetime metric and for the Maxwell equations themselves. In
\secref{numerics} we describe the numerical methods used to solve the
equations. We summarize the numerical results and discuss the
implications for future work in \secref{results}. Additional detail is
provided in the Appendix.

\section{Formalism}
\label{sec:formalism}
\subsection{Spacetime Metric}
\label{sec:metric}
As in Ref.~\cite{misner:deserFest} 
the background metric for this study will be the de Sitter spacetime
\begin{equation} 
ds^2 = -dT^2 +dX^2 + dY^2 + dZ^2 + (R^2/L^2)(dT-dR)^2  
\label{eq:deSorig} 
\end{equation} 
with an artificial cosmological constant $\Lambda = +3/L^2$. We are
interested in the limit $1/L^2 \rightarrow 0$ when this becomes Minkowski
spacetime, and the cosmological constant is used to make boundary
conditions numerically simpler (we hope) at the de Sitter horizon than
they would be at flat spacetime's $\scri^+$. The de Sitter horizon is the null
hypersurface $R=L$ in this metric.  We introduce the coordinate
changes
\begin{subequations}
\begin{eqnarray}
  \frac{T}{s} & = & u + \frac{r^2/2}{1-r^2/4}
\label{eq:HsliceU} \\
  \frac{X^i}{s} & = & \frac{x^i}{1-r^2/4}
\end{eqnarray}
\label{eq:AnMR}
\end{subequations}
as used in Refs.~\cite{MSL:KITP,misner:deserFest},
where $R^2 = X^2 +Y^2 +Z^2 \equiv X^i X^i$, $r^2 = x^2+y^2+z^2 \equiv
x^i x^i$, and $s$ is a constant scaling factor whose geometrical 
significance
can be read from Eq.~(\ref{eq:hyperboloid}) below.
This brings $\scri^+$ in to $r=2$ in the Minkowski case and in the
de Sitter case makes this a spacelike hypersurface beyond the de Sitter
horizon which is part of de Sitter spacetime's future causal boundary  
which is also called $\scri^+$.
The hypersurfaces of constant $u$ are then hyperboloids 
\begin{equation}
\label{eq:hyperboloid}
     [T-s(u-1)]^2 - R^2 = s^2
\end{equation}
in Minkowski spacetime and are also asymptotically null spacelike
hypersurfaces in the de Sitter modification.
These hyperboloids in Minkowski spacetime have constant curvature
of radius $s$ which is therefore the scale of the region in which they
are approximately flat. Also the time interval $\Delta T$
at the spatial origin between the $u$ slice on which a light signal is
emitted there and the $u$ slice on which it reaches $\scri+$ is
$\Delta u = 1$ or $\Delta T = s$.  Thus $s$ may also be called
the scri-delay.
This coordinate change leads to a metric which is singular at $r=2$, but 
only in a
conformal factor $s^2/q^2=s^2/(1 - r^2/4)^2$ which
does not appear in the Maxwell equations. 
Thus we can drop this conformal factor and our test problem is 
to solve the Maxwell equations in the resultant metric which
is of the form
\begin{equation}
       ds^2 = - \alpha^2 dt^2 
              + \gamma_{ij}(dx^i + \beta^i dt) (dx^j + \beta^j dt)
\label{eq:g3+1}
\end{equation}
with $t=u$. 
For use in later modifications we include a function $W(r)$ in the 
definition
of the metric functions in (\ref{eq:g3+1}).
For now we choose $W=0$ and then the following equations give the
analytically continued and conformally regulated de Sitter metric
described
above:
\begin{subequations}
\begin{eqnarray}
\alpha & = & 
\frac{(1-W)(1+\frac{r^2}{4})}{\sqrt{1+\left(\frac{sS}{L}\right)^2}} \\
\beta^i & = & 
-x^i \frac{1-W + \left(\frac{s}{L}\right)^2 S}{1+\left(\frac{sS}{L}\right)^2}\\
\gamma_{ij} & = & \left(\frac{sS}{Lr}\right)^2 x_i x_j + \delta_{ij}
\end{eqnarray}
\label{eq:metric}
\end{subequations}
where
\begin{equation}
S = \frac{(1-W)r}{(1+\frac{r}{2})^2} 
+ \frac{1}{4} \left(\frac{L}{s}\right)^2 W.
\label{eq:S}
\end{equation}

To see the behavior of the light cones for this regulated metric, we
calculate the coordinate speed of light $dr/du$ in radial null
directions.
This gives
\begin{equation}
\label{eq:lightspeed}
  v = \frac{r + r (s^2/L^2) S \pm (1+\frac{r^2}{4})}{1 + (s^2/L^2) S^2}
    \quad .
\end{equation}
For the de~Sitter values above this reduces to
\begin{equation}
\label{eq:vout}
   v_{\rm out} = (1+ \half r)^2
\end{equation}
and
\begin{equation}
\label{eq:vin}
  v_{\rm in} = \frac{-(1-\half r)^2 + (s^2/L^2)(1+\half r)^2 S^2}%
    {1 + (s^2/L^2) S^2}
  \quad .
\end{equation}
Note that the constant $r$ hypersurfaces are never spacelike
when $s/L=0$ since then the inward speed of light is $v_{\rm in} \leq
0$.
When the de~Sitter option is used one finds $v_{\rm in} > 0$, i.e., both
the inward and the outward sides of the lightcone point toward
increasing $r$, for an interval around $r=2$.

In evolution codes which permit setting the computational boundary on a
coordinate sphere (such as those described in
\cite{kls05:boundary,thornburg:boundary,lsu:lrt,lsu:ddst}) the outer
boundary could be set at $r=2$ or slightly beyond in the metric
described above (or ones asymptotically similar).
However, if a cubic boundary enclosing $\scri^+$ is to be used, there
are additional behaviors beyond $\scri^+$ that need attention.
For example, if the artificial de~Sitter horizon is set at $L = 10 s$,
then the $r=\text{constant}$ hypersurfaces are spacelike only in the
narrow interval of (approximately) $r=1.81$ to $r=2.21$.
But a boundary cube enclosing $r=2$ has corners at $r=2 \sqrt{3} \approx 
3.46$.
Thus on most of such a cubic boundary the light cones (from Eq.~(\ref{eq:vin})) 
would allow inwardly propagating waves, requiring careful
attention to the boundary conditions which one hopes to avoid.

Fig.~\ref{fig:S-lightcone_deSitter}  
below shows the coordinate speeds for light rays on the
inward side of the light cone $v_\text{in}$, on the outward side of the
light cone $v_\text{out}$, and for a timelike center of the light cone
$v_\text{center} = -\beta$ normal to the time slices of constant $u$.
\begin{figure}[h]
\rotatebox{0}{\epsfig{file=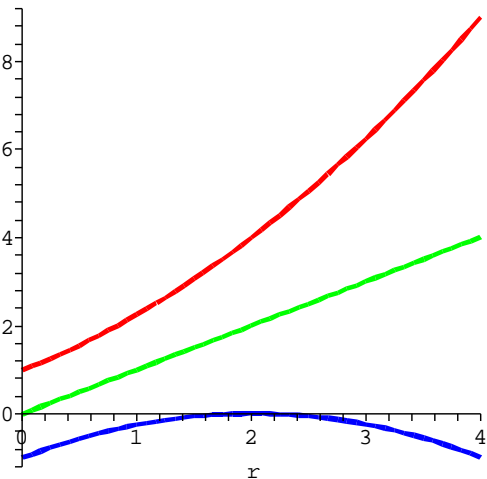,width=3in}}
\caption{Coordinate speeds for the de Sitter lightcone: inward
($v_\text{in}$, blue, lowest curve), normal to the spacelike time slices 
($-\beta$, green, middle curve), and
outward ($v_\text{out}$, red, highest curve). For the Minkowski metric the graph looks very
similar, except that $v_\text{in}$ is there never positive.}
\label{fig:S-lightcone_deSitter}
\end{figure}
This allows light rays to move inward ($v_\text{in}<0$) in 
some of the region beyond $\scri^+$ within a cubical computational grid.

In this case we suggest that the metric be modified outside $r=2$ so that
all $r=\text{constant}$ hypersurfaces beyond $r=2$ are spacelike ---
i.e., so that $v_{\rm in} > 0$ in the entire computational domain beyond
$r=2$.
Analytically this would have no effect on the physical solution inside
$r=2$ provided all fields propagate causally.
This modification is also straightforward to do in the test case of a Maxwell field
with a fixed background metric.
If these methods were to be developed for the Einstein equations
(without using a spherical boundary such as $r=2$) it would necessitate
distinguishing, in the Einstein equations beyond $r=2$, the $g_{\mu\nu}$
metric being evolved from the appearances of the metric as coefficients
of the principle derivatives which determine the causal relations in the
evolution algorithm.

In the Minkowski case $s/L=0$ (no artificial cosmology) it is not
possible to make a $C^2$ change in the metric only beyond $r=2$ which
avoids inward pointing null rays just beyond $r=2$ --- in that case one
has, from the $r \leq 2$ formula above for $v_{\rm in}$ that
$(d^2/dr^2)v_{\rm in}= -\half$ at $r=2$ so that $v_{\rm in}$ must have a
local
maximum value of zero at that point and is necessarily negative (inward)
in a neighborhood of that point, including some $r>2$ points.
For the de~Sitter case there is again a maximum of $v_{\rm in}$ near
$r=2$,
but its value there is positive so it can be smoothly turned upward (to
avoid zero or negative values) in suitable intervals beyond the
de~Sitter horizon (and beyond $r=2$ for small values of $s/L$).

In Eqs.~(\ref{eq:metric})~and~(\ref{eq:S}), assuming $s/L>0$, we choose to take
\begin{equation}
\label{eq:modify}
W(r) = \Theta(r-2) 
\frac{(r-2)^4}{\left[\left(r-2\right)^2+\frac{1}{2}\right]^2}
\end{equation}
to be a smooth step function that modifies the metric beyond $r=2$ in a way
that keeps the ``incoming'' coordinate speed of light 
\(v_{in} = dr/du\) positive (that is outgoing) when $r>2$.  The function
$W$ is a $C^3$ modification of \(\Theta(r-2)\), the Heaviside step function
that jumps from 0 to 1 at $r=2$.  
Fig.~\ref{fig:S-lightcone} below shows the modified light cone directions
in this case.

Note that for \(r \leq 2\), where $W=0$, this metric
has the surprising property that the outgoing speed of light
\(v_{out} = (1+r/2)^2\) is independent of the cosmological constant
parameter $L$. 
Although in general the presence of a cosmological constant alters the
causal structure of a spacetime, and indeed it does modify the ingoing
radial speed of light in the present example, it does not change the
outgoing radial speed of light from that of the original Minkowski spacetime.
This fact will prove useful for numerical diagnostics.  

\begin{figure}[h]
\rotatebox{0}{\epsfig{file=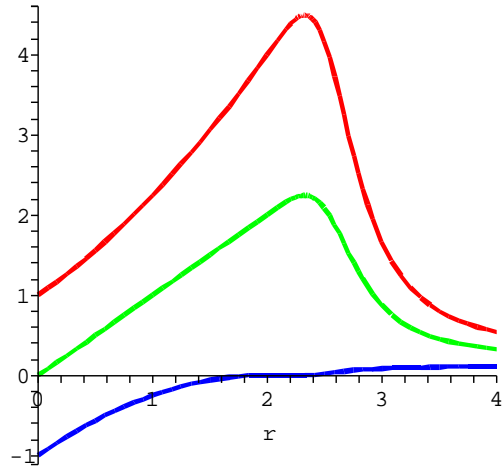,width=3in}}
\caption{Coordinate speeds for the modified de Sitter
(eqn.~\ref{eq:modify}) lightcone: inward light rays, ($v_\text{in}$, blue, lowest curve),
motion normal to the spacelike time slices ($-\beta$, green, middle curve), and outward light
rays ($v_\text{out}$, red, highest curve).}
\label{fig:S-lightcone}
\end{figure}

\begin{figure}
\rotatebox{0}{\epsfig{file=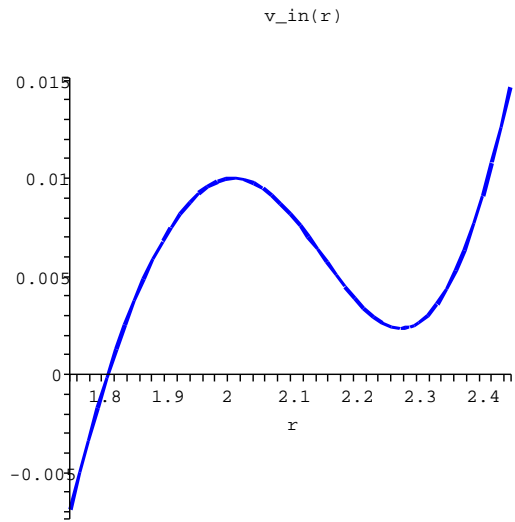,width=3in}}
\caption{Coordinate speeds inward ($v_\text{in}$) for light rays 
in the modified (eqn.~\ref{eq:modify}) de Sitter metric.}
\label{fig:S-v_in}
\end{figure}

\begin{figure}[h]
\rotatebox{0}{\epsfig{file=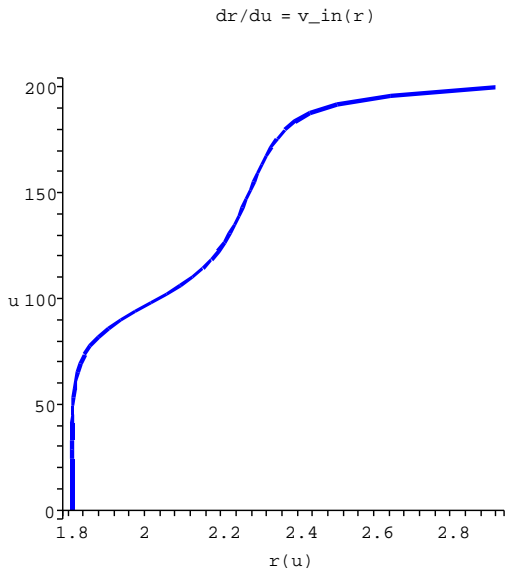,width=3in}}
\caption{World line of a light ray on the inward side of the light cone 
in the modified de Sitter metric.}
\label{fig:S-inward_ray_longterm}
\end{figure}

This metric surgery has an additional side effect worth noting.
For this modified de Sitter metric using Eq.~(\ref{eq:modify}) one finds
very low inward light velocities in the region just outside $\scri^+$ as 
seen in Fig.~\ref{fig:S-v_in}.
This can lead to fields propagating along the inward side of the lightcone 
spending very long periods of time in the region $r\approx 2.3$ as illustrated
by the world line of a light ray in Fig.~\ref{fig:S-inward_ray_longterm}.

\subsection{Maxwell Equations}
\label{sec:maxwell}
We take the Maxwell equations in the form given in the DeserFest paper
\cite{misner:deserFest}.
The quantities
involved are all various components of the usual 4-dimensional Maxwell
fields $F_{\mu\nu}$ and $\mathcal{F}^{\mu\nu} =
\sqrt{-g}g^{\mu\alpha}g^{\nu\beta} F_{\alpha\beta}$.
In particular, for any $3+1$ metric of the form 
Eq.~(\ref{eq:g3+1}) we define,
\begin{subequations}
\begin{equation}
  \label{eq:3+1Fcomp}
     \mathcal{F}^{0i} = \mathcal{D}^i \quad , 
     \quad  \mathcal{F}^{ij} = [ijk] H_k
\end{equation}
and 
\begin{equation}
F_{i0} = E_i  \quad , \quad F_{ij} = [ijk] \mathcal{B}^k
              \quad .
\end{equation}
\end{subequations}
Here $[ijk]$ is the completely antisymmetric symbol $[ijk] = 0, \pm
1$ with $[123] = +1$.
The fundamental fields in the formulation are $\mathcal{B}^i$
and $\mathcal{D}^i$. The constraints read
\begin{equation}
  \partial_i \mathcal{B}^i = 0 = \partial_i \mathcal{D}^i 
  \quad .
  \label{eq:constr}
\end{equation}
We need only take regular
partial (rather than covariant) derivatives in our formulation of 
Maxwell's equations because of judicious use of tensor densities
rather than tensors (cf. Exercises 22.8 and 22.9 in \cite{MTW}).
The resulting fields in Eqs.~(\ref{eq:constr}) are each three dimensional 
vector densities so this
equation is three dimensionally covariant. In terms of these
primary fields, two auxiliary fields ${E}_i$ and
${H}_i$ are computed by the formulas
\begin{subequations}                                        
\label{eq:defEH}
\begin{equation}
\label{eq:defE}
  {E}_i = (\alpha/\sqrt{\gamma})\gamma_{ij}\mathcal{D}^j
    + [ijk]\beta^j \mathcal{B}^k 
\end{equation}
and
\begin{equation}
\label{eq:defH}
  {H}_i = (\alpha/\sqrt{\gamma})\gamma_{ij}\mathcal{B}^j 
     - [ijk]\beta^j \mathcal{D}^k
      \quad .
\end{equation}
\end{subequations}                                        
The evolution equations are then
\begin{subequations}
\label{eq:EBevol}
\begin{equation}
\label{eq:Bevol}
   \partial_0 \mathcal{B}^i  = - [ijk]\partial_j E_k
\end{equation}
and
\begin{equation}
\label{eq:Eevol}
   \partial_0 \mathcal{D}^i = [ijk]\partial_j H_k
   \quad .
\end{equation}
\end{subequations}
Note that the metric does not appear in Eq.~(\ref{eq:constr}) or
Eqs.~(\ref{eq:EBevol}), and that the metric quantities which do appear in
Eqs.~(\ref{eq:defEH}) are conformally invariant. 
This is a first order partial differential equation system in flux conservation form.

\subsection{Initial Data}
For initial data and some testing purposes we have used the solution of
the flat spacetime Maxwell equations employed by 
Knapp et al.\ \cite{baumgarte:em} and subsequently by 
Fiske \cite{fiske:constem}.  This solution is a wave pulse, which
propagates smoothly from $\scri^-$ through the origin and out to 
$\scri^+$.  It consists of a toroidal $\mathcal{D}$ field and a
poloidal $\mathcal{B}$ field generated from a vector potential,
described in Ref.~\cite{misner:deserFest},
\begin{equation}
\label{eq:A-example}
A = A_i dX^i = f \sin^2 \theta d\phi
\end{equation}
where
\begin{equation}
\sin^2 \theta d\phi = \frac{1}{R^2} \left(X dY = Y dX \right)
= \frac{1}{r^2} \left(x dy - y dx \right)
\end{equation}
and
\begin{equation}
\label{eq:f_init-example}
f = \left(\frac{1}{R} - 2\lambda U \right) e^{-\lambda U^2}
- \left(\frac{1}{R} + 2\lambda V\right) e^{-\lambda V^2}
\end{equation}
where the constant $\lambda$ parameterizes the pulse width.  
In these equations \(U \equiv T - R\) and \(V \equiv T + R\) are the
Minkowski retarded and advanced times, which are then expressed in terms
of the compactified coordinates $u$ and $r$ as $U/s=u-2+2/(1+\half r)$ and
$V/s=u-2+2/(1-\half r)$. Finally, $\mathcal{B}^i$ and $\mathcal{D}^i$ are
calculated by differentiating this $A$ field. 
The term containing $e^{-\lambda V^2}$ can be smoothly omitted when
$r>2$ since all its derivatives vanish at $r=2$, but the term with
$e^{-\lambda U^2}$ must be retained to preserve the constraints, and we
thus have nonzero initial data in the unphysical region $r>2$, i.e. beyond
$\scri^+$, where initial data are not physically meaningful. 
As the initial pulse is centered at $r=0$, we would ideally like to make it sufficiently narrow 
so that its amplitude would rapidly taper off and become negligible beyond $r>2$.
However, we did not achieve enough resolution to reduce the pulse 
width to make these unphysical data beyond $r>2$ insignificant.

For all of the simulations described in this paper, we choose 
$\lambda = 1$ (or, equivalently, we measured distances in the uncompactified
space in units of $\lambda^{-1/2}$).  The pulse width is then 
$\lambda^{-1/2}$.  We also always choose $s=1$ to set the size of the 
region near the origin where the hyperboloidal slices are somewhat flat.
Our results reported below used either $s/L=0$ for the Minkowski case, 
or $s/L=0.1$ for the de Sitter and modified $\lbrack$Eq.~\ref{eq:modify}$\rbrack$ de Sitter cases. 
Thus the $r=\text{constant}$ hypersurfaces were not spacelike for the de Sitter case 
when $r>2.21$

\section{Numerics}
\label{sec:numerics}
A key point in our approach to this problem is that we do not
require special numerical techniques to handle the compactified spacetime
at or within $\scri^+$. 
At all interior points, we use second-order, finite-difference approximations
to all derivatives.  At the computational boundary beyond $\scri^+$, 
we use a second-order, one-sided stencil for normal derivatives.
We integrate forward in time using the iterated
Crank-Nicholson method commonly used in numerical relativity 
codes \cite{teuk:icn}.

Our decision to use one-sided differencing at the computational boundary
was (in our view) the simplest thing to do. It turns out, however, to be
equivalent to the black hole excision boundary conditions first studied
in Ref.~\cite{maya:moving}. To be precise, along a normal to the
boundary, the composition of third order extrapolation to a cell just
outside the boundary (as used for excision) with second-order center
differencing at a cell just inside the boundary gives the same stencil
as second-order one-sided differencing at the cell just inside the
boundary. This is consistent with our mental picture of excising a
region exterior to $\scri^+$, and reflects the similarity between
$\scri^+$ (at which all characteristics point outward) and an apparent
horizon of a black hole (at which all characteristics point inward). As
with black hole excision, it is critical for stable and accurate
simulations that all light cones at the excision boundary tilt towards
the excised region, hence the need for the beyond-$\scri^+$ metric
modification described in Sec.~\ref{sec:metric}.  Without these 
Eq.~(\ref{eq:modify}) modifications, the light cones would point outward only 
in a neighborhood of $\scri^+$ which did not extend out to the 
computational boundary of the cubic grid enclosing $\scri^+$.

\begin{figure}
\rotatebox{-90}{\epsfig{file=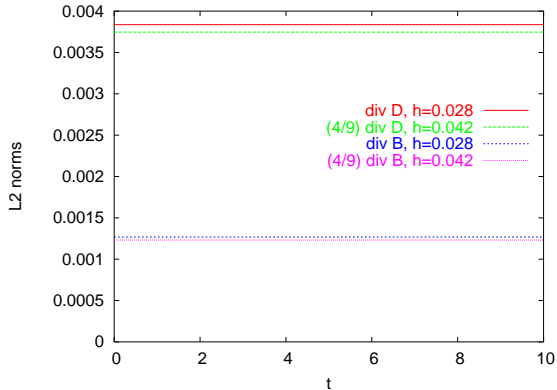,height=3in}}
\caption{The L2 norms (over the region $r\leq 2$) of the constraints
plotted vs.\ time. The higher resolution data is multiplied by the factor
appropriate to demonstrate second order convergence.
The initial values arise from $\mathcal{B}$ and $\mathcal{D}$ fields
which are exact analytic solutions of the constraints, evaluated at the
grid points and then differenced to form the constraints.
}
\label{fig:convergence}
\end{figure}

We have verified that our code converges at second order accuracy
throughout the physical domain. We paid special attention to the region in the
vicinity of $\scri^+$, and we are satisfied that the code converges there
as well. Fig.~\ref{fig:convergence} explicitly shows a two-point
convergence test on the constraints over the physical region $r \leq 2$,
which are seen to be second-order convergent to zero and nearly
constant.

\section{Results}
\label{sec:results}
When evolving with the original, conformal, de Sitter metric given in 
Eq.~(\ref{eq:deSorig}), large errors eventually appear
near the computational grid boundary, grow exponentially in time, and 
propagate inside $\scri^+$.
This problem was slightly worse when the cosmological constant was omitted.  
These cataclysmic errors originated at the grid boundary which in these cases 
was not a spacelike hypersurface for which excisions (at apparent horizons) were 
designed.
As anticipated in Sec.~\ref{sec:metric} we proceeded to artificially
tilt the lightcones outwards beyond $\scri^+$ to make the grid boundaries
spacelike.

\begin{figure}
\epsfig{file=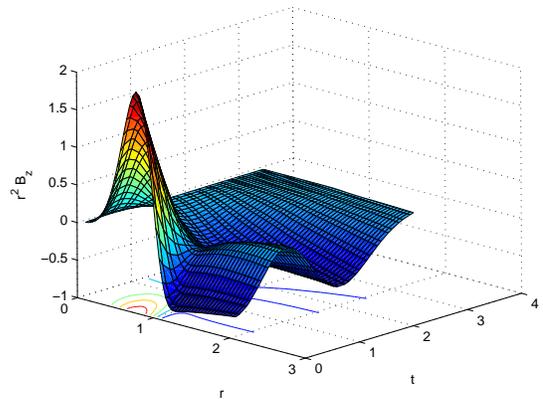,width=3in}
\caption{A spacetime graph of $r^2 \mathcal{B}^z$. Notice that the wave
passes through $\scri^+$, located at $r=2$ in our coordinate system.}
\label{fig:Bz3D}
\end{figure}

\begin{figure}
\rotatebox{-90}{\epsfig{file=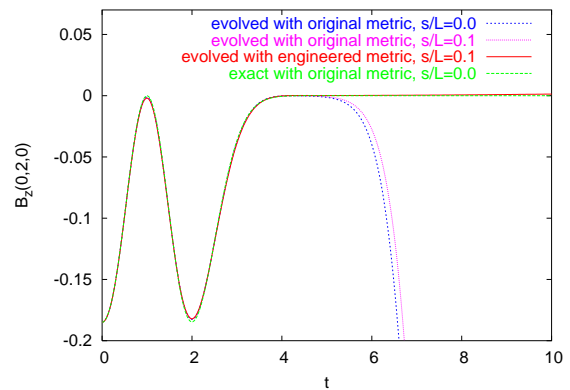,height=3in}}
\caption{The time evolution of $\mathcal{B}^z$ at a point on $\scri^+$.
We show both a numerical solution generated by our code using the
modified de Sitter (engineered lightcones beyond $\scri^+$) and the
analytic solution with zero cosmological constant. The close agreement
of the lines suggests that we are able to accurately capture the
behavior of the fields as they pass through $\scri^+$. Other numerical
evolutions based on the original (compactified, analytically continued)
Minkowski metric ($s/L=0.0$) and the similar de Sitter metric
($s/L=0.1$), for which the grid boundaries were not spacelike, fail after
$t \equiv u = 4$.}
\label{fig:Bz1D}
\end{figure}

Fig.~\ref{fig:Bz3D} shows a spacetime view of the $z$ component of the
magnetic field for an evolution using the modified (eqn.~\ref{eq:modify})
de Sitter metric.
In order to make clear that the wave is cleanly passing through $\scri^+$
through the course of the evolution, we have multiplied the field by
$r^2$, making the scale of the field comparable at the origin and at 
$\scri^+$.  This can be seen in a more local way in Fig.~\ref{fig:Bz1D},
which shows the time evolution of $\mathcal{B}^z$ at a single point on
$\scri^+$, for four cases. 
One is the analytic value of the field (in extended Minkowski spacetime)
for comparison. 
The two cases which failed to maintain vanishing fields in the wake of
the outgoing pulse correspond to the two (Minkowski and de Sitter) metrics
for which the modifications needed to make the grid boundaries beyond
$\scri^+$ spacelike $\lbrack$Eq.~(\ref{eq:modify})$\rbrack$ were not implemented.

Note that the analytic solution is with zero cosmological constant, whereas
the numerically evolved solution is with finite cosmological constant, and yet
the waveforms coincide.  This phenomenon can be understood as follows.
Because, as pointed out in Sec.~\ref{sec:metric}, the outgoing speed of light is 
the same in either case, the phase of outgoing waves can be
said to be independent of a cosmological constant here.  Further,
since by construction the initial wave pulse is identical in our Minkowski and
de Sitter simulations, by conservation of energy the wave amplitudes at 
$\scri^+$ are expected to be commensurate.

These results show that many desired aims of this approach are achieved.
Waves propagate smoothly through $\scri^+$ and no difficulties appear to
arise in the neighborhood of $\scri^+$ The compactified hyperboloidal
slices allow waves to appear at $\scri^+$ a very short computational
time after they originate in the central region. No modifications were
needed to integration algorithms designed for conventional spacelike
Cauchy slicings. The excision at a cubic computational boundary,
however, only maintained acceptable behavior for a time several times
the pulse width, and ultimately led to behaviors in the unphysical
region beyond $\scri^+$ which became intolerable. An example is given
in Fig~\ref{fig:S-outerconstraints} where the constraint violations
beyond $\scri^+$ are seen eventually to increase exponentially. In
addition, by using analytically constructed initial values satisfying
the constraints, we had to accept nonzero initial conditions in the
unphysical region $r>2$ beyond $\scri^+$, and the subsequent evolution of
these fields complicated the computation beyond $\scri^+$.

We believe that further developments of Cauchy evolutions with
compactified hyperboloidal slicings and artificial cosmology should be
done with the computational/excision boundary a spherical hypersurface
at or modestly beyond $\scri^+$. 
The means for doing this have been developed, e.g. 
\cite{kls05:boundary,thornburg:boundary,lsu:lrt,lsu:ddst}.
Then the possibility remains open that well posed wave equations could
run indefinitely. 
As explained in the supplementary material (Appendix), here
we did not use the differential equations from Sec.~VI of
\cite{misner:dorothy} as our lightcone engineering (see
Sec.~\ref{sec:metric}) beyond $\scri^+$ makes the constraints there
increase exponentially even from analytic arguments. 
A spherical boundary could be spacelike at $r>2$ without the need for
such artificial tilting of the lightcones, and thus allow the use of the
formulation \cite[Equations 26]{misner:dorothy} where constraints evolve
causally.

\acknowledgments{This work was supported in part by NASA Space Sciences
Grant ATP02-0043-0056.  JvM and DRF were also supported in part by
the Research Associateship Programs Office of the National Research
Council.  CWM gratefully acknowledges the hospitality of the AEI, where
much of this work was developed.}

%

\bibliography{Misner.bib,gr.bib}

\clearpage

\appendix

\setcounter{figure}{0}
\renewcommand{\thefigure}{\thesection\arabic{figure}}

\section{Auxiliary material}
\label{sec:Append}

\begin{figure*}[!t]
\rotatebox{-90}{\epsfig{file=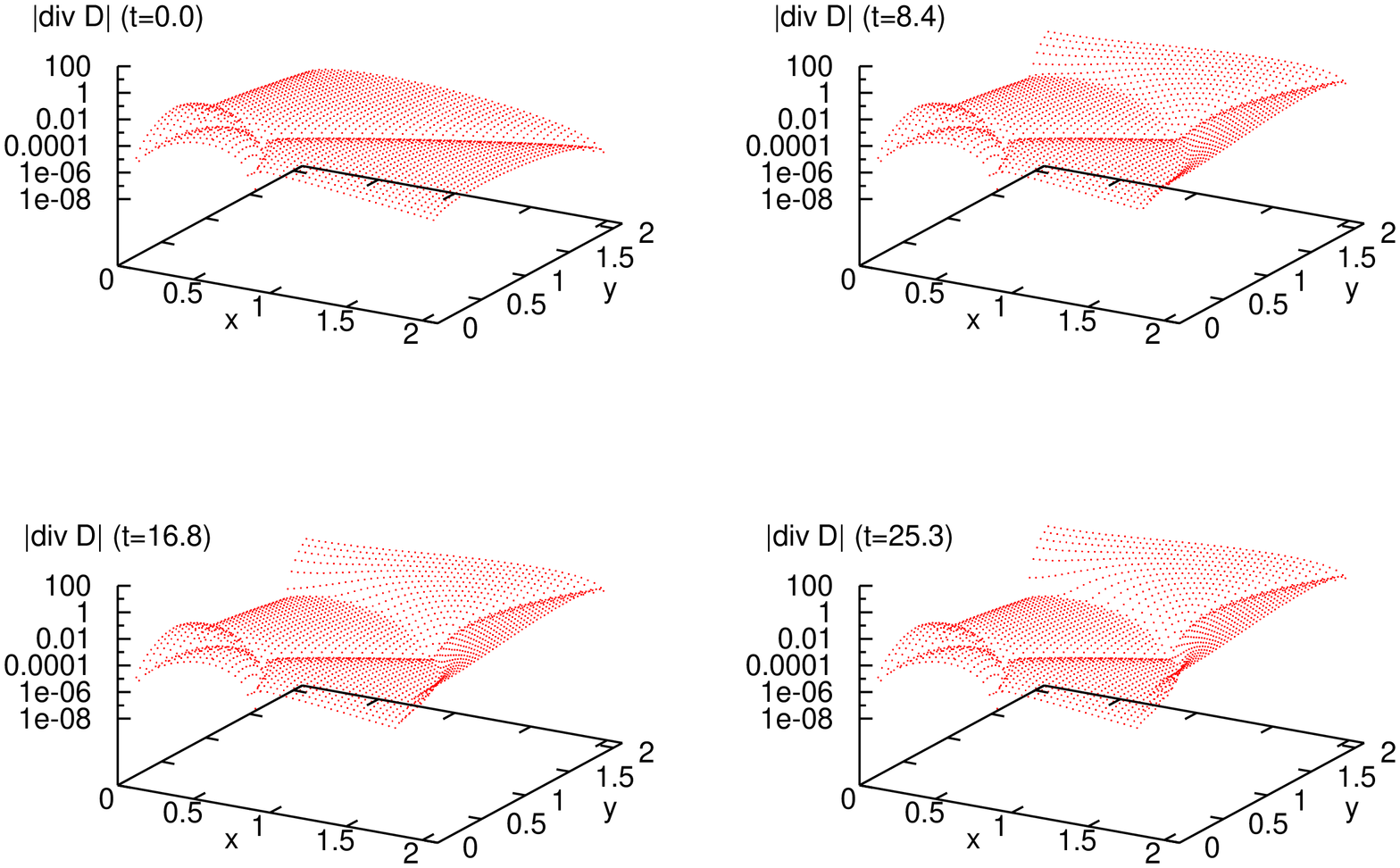,height=6in}}
\caption{Samples of the time evolution of the constraint violation
$|\text{div} \mathcal{D}|$ at $z=0$ within our comptational 
cube $|x^i|<2.1$ whose edges are at $r=2.97$ .}
\label{fig:S-divD}
\end{figure*}

This section is mainly a collection of graphs which illustrate
properties that are merely asserted in the main article.


\subsection{Constraints}

\begin{figure}
\rotatebox{-90}{\epsfig{file=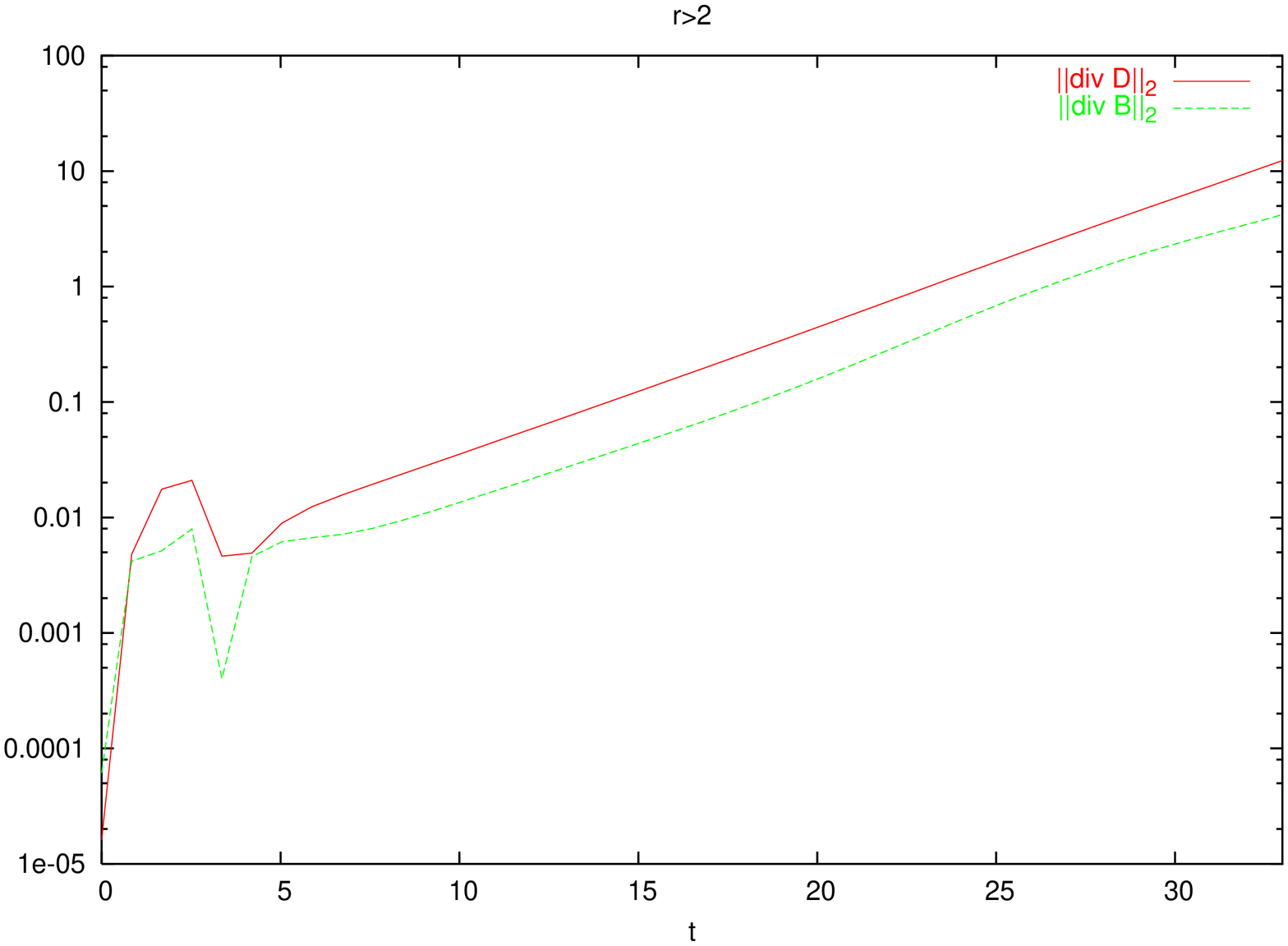,height=3in}}
\caption{The time evolution of an L2 norm of the constraint violations
over the unphysical region $r>2$ beyond $\scri^+$ within our computational 
cube $|x^i|<2.1$ whose corners are at $r=3.64$ .}
\label{fig:S-outerconstraints}
\end{figure}

Even when the modified de Sitter metric is used, the solution in the
unphysical region deteriorates as shown in
Figs.~\ref{fig:S-divD} and \ref{fig:S-outerconstraints}.

\begin{figure}[p]
\rotatebox{-90}{\epsfig{file=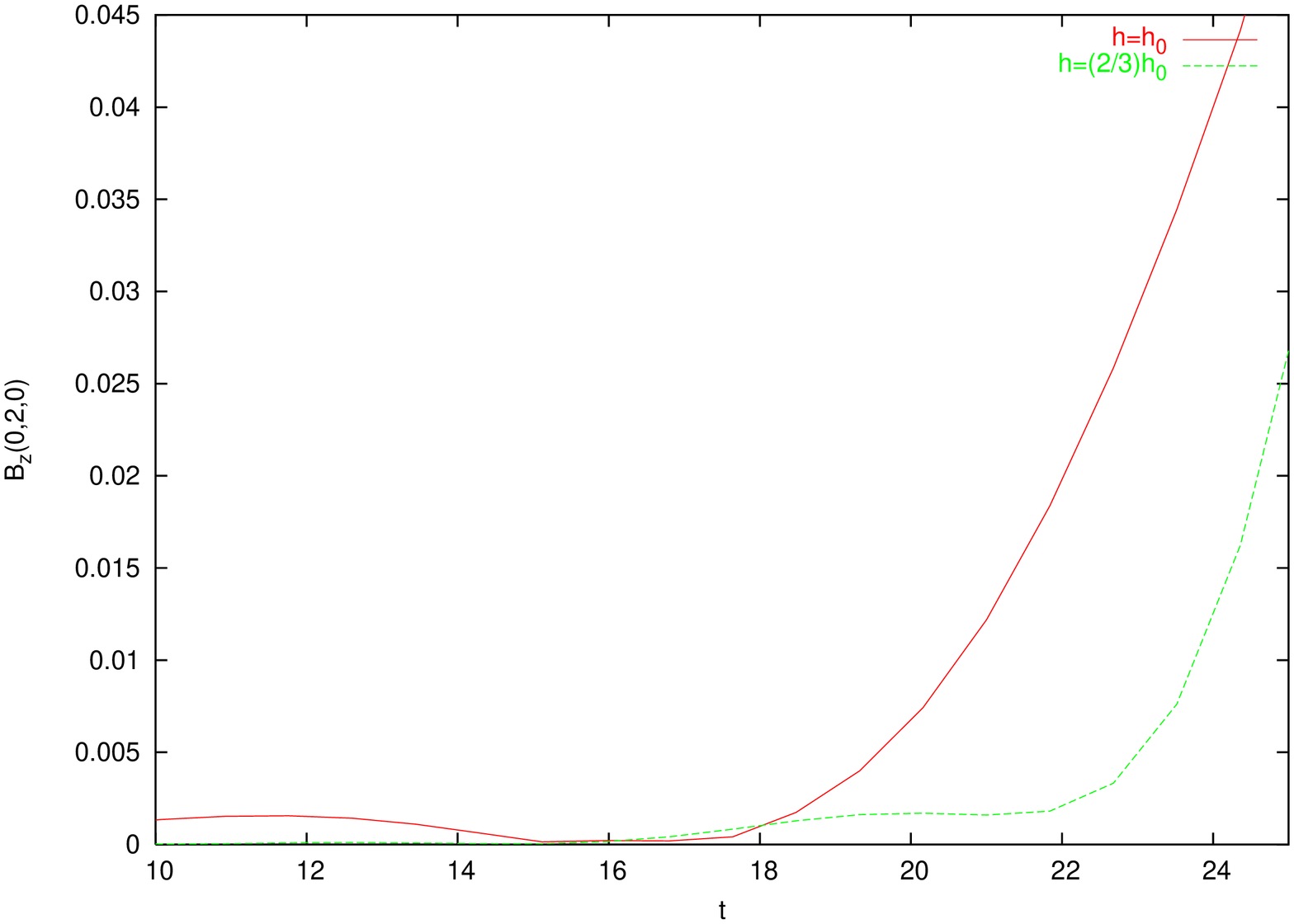,height=3in}}
\caption{A continuation of the test wave pulse to later times at $\scri^+$.}
\label{fig:S-conv_wave}
\end{figure}

In spite of these problems, the field at $\scri^+$ remains zero after
the wave pulse passes (Fig.~\ref{fig:Bz1D}) for a moderately long time,
which can be lengthened by using higher resolution. See
Fig.~\ref{fig:S-conv_wave}.
Better control of the constraints should be possible using the equation
\begin{figure}[t]
\rotatebox{0}{\epsfig{file=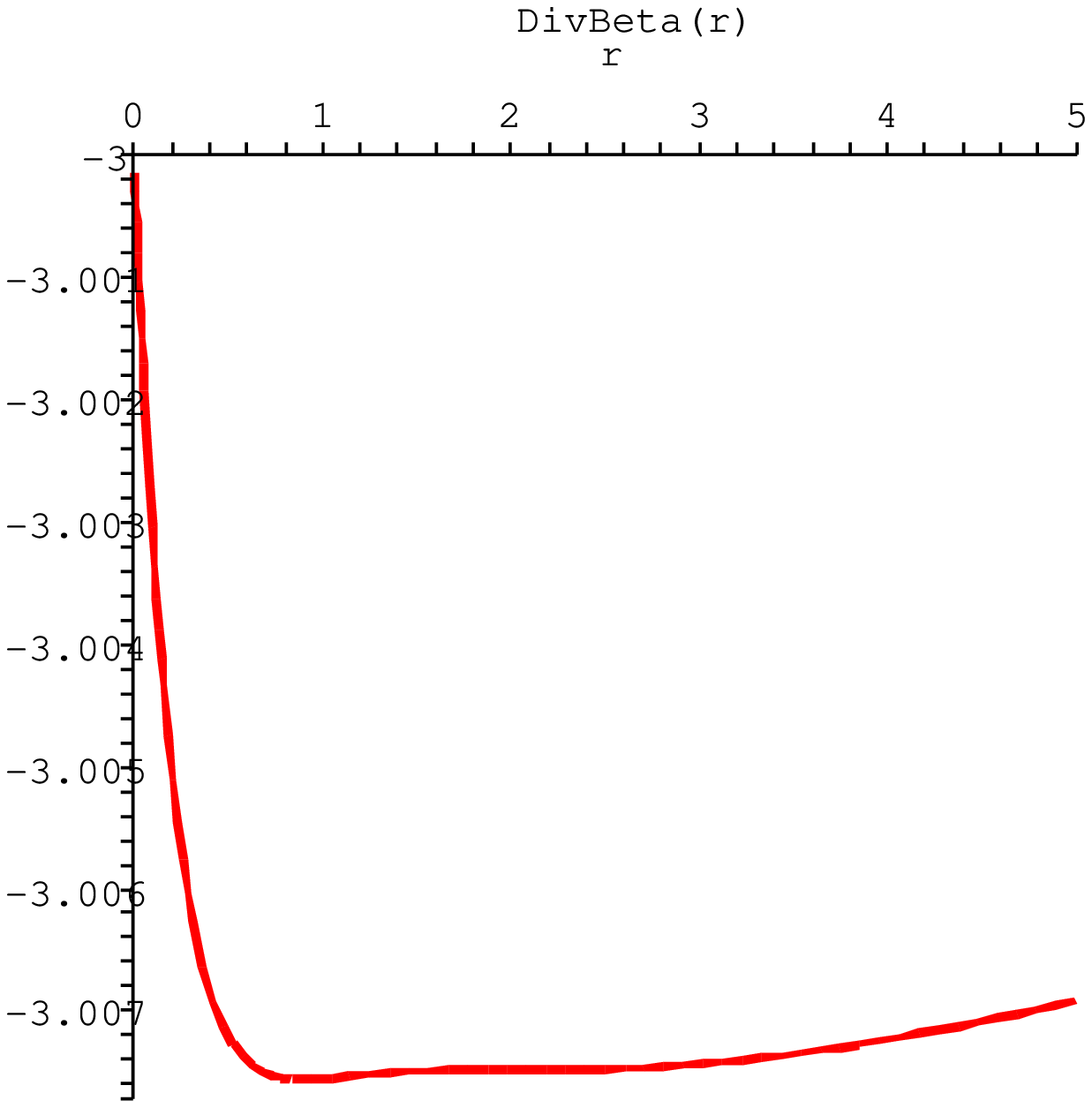,width=3in}}
\caption{The damping factor for constraints in the de Sitter metric.}
\label{fig:S-divBeta_deSitter}
\end{figure}
\begin{figure}
\rotatebox{0}{\epsfig{file=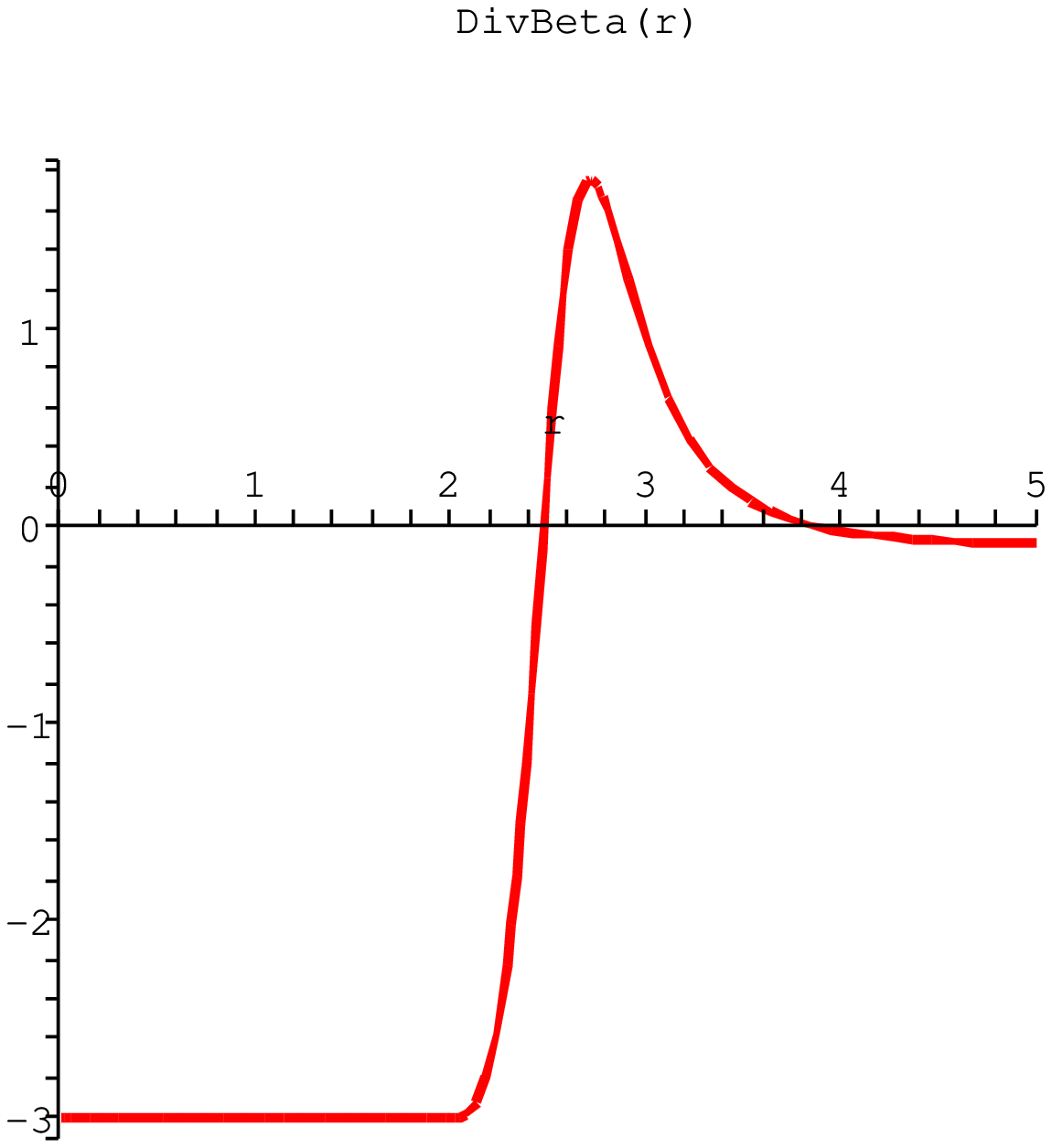,width=3in}}
\caption{The damping factor for constraints in the modified de Sitter metric.
Where $\text{div} \beta$ is positive, the constraints should grow exponentially
at the plotted e-folding rate.}
\label{fig:S-divBeta}
\end{figure}
system \cite[Equations 26]{misner:dorothy} where the constraints
propagate inside the lightcone and damp at a rate $-\partial_i \beta^i$.
However this is damping only for the Minkowski or de Sitter metrics as
seen in Fig.~\ref{fig:S-divBeta_deSitter} and becomes anti-damping for
our modification of the de Sitter metric (see Fig.~\ref{fig:S-divBeta} )
which makes our cubic grid boundaries spacelike.

\subsection{Initial conditions beyond $\scri^+$.}
\label{sec:ICbeyond}

A second reason which makes the use of the full cubic grid (rather than
only a region inside a spherical boundary such as $r=2$ or $r=2.1$)
inappropriate for numerical computation is the need for initial values
beyond $\scri^+$, that is beyond $r=2$, which is $R=\infty$. 
\begin{figure}[t]
\rotatebox{0}{\epsfig{file=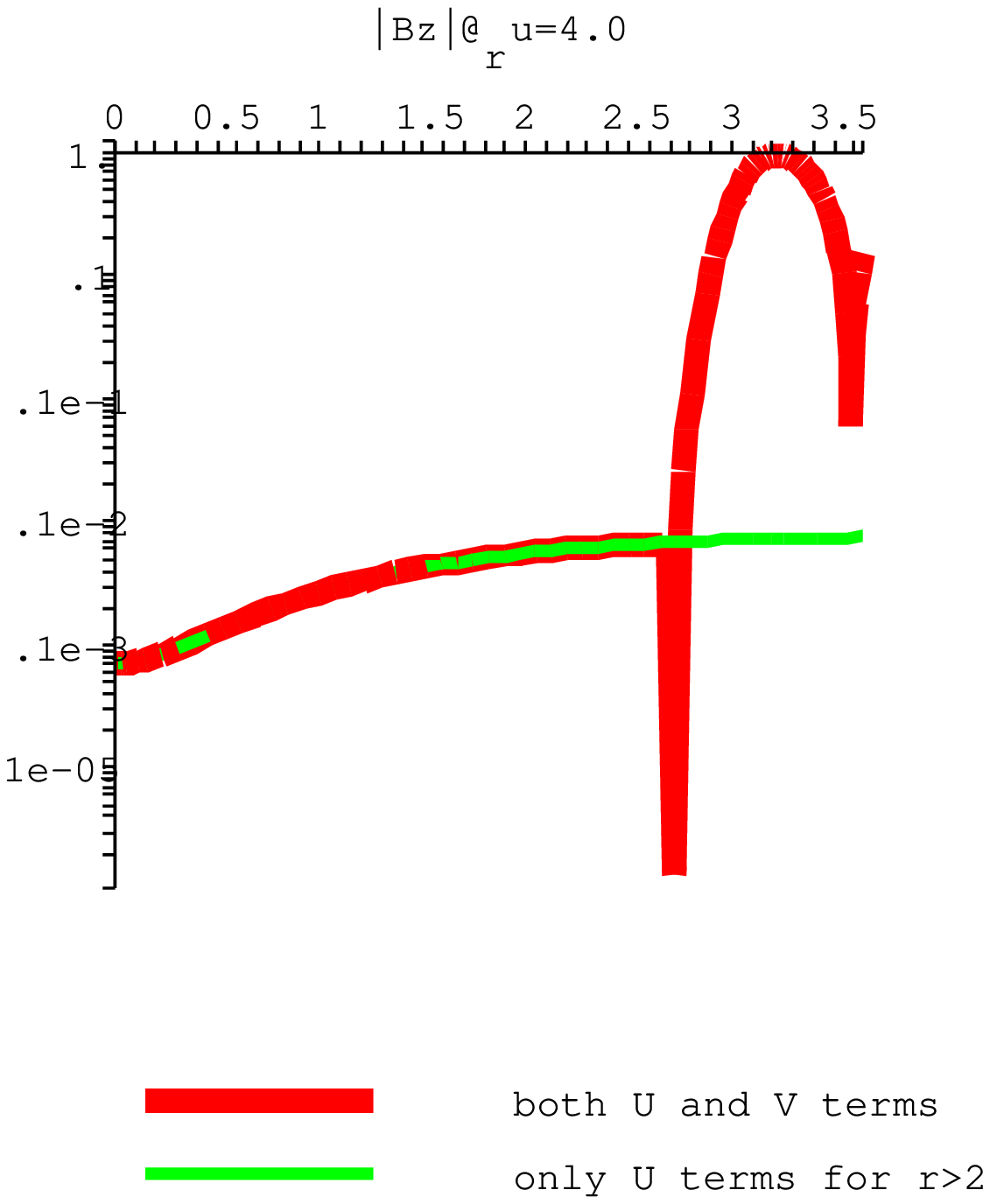,width=3in}}
\caption{\label{fig:S-exact_log_incoming40}For the analytic 
solution in Minkowski spacetime: The magnetic
field at $z=0$ after the pulse from Fig~\ref{fig:Bz1D} has 
exited the computational grid.}
\end{figure}

\begin{figure*}[t]
\rotatebox{-90}{\epsfig{file=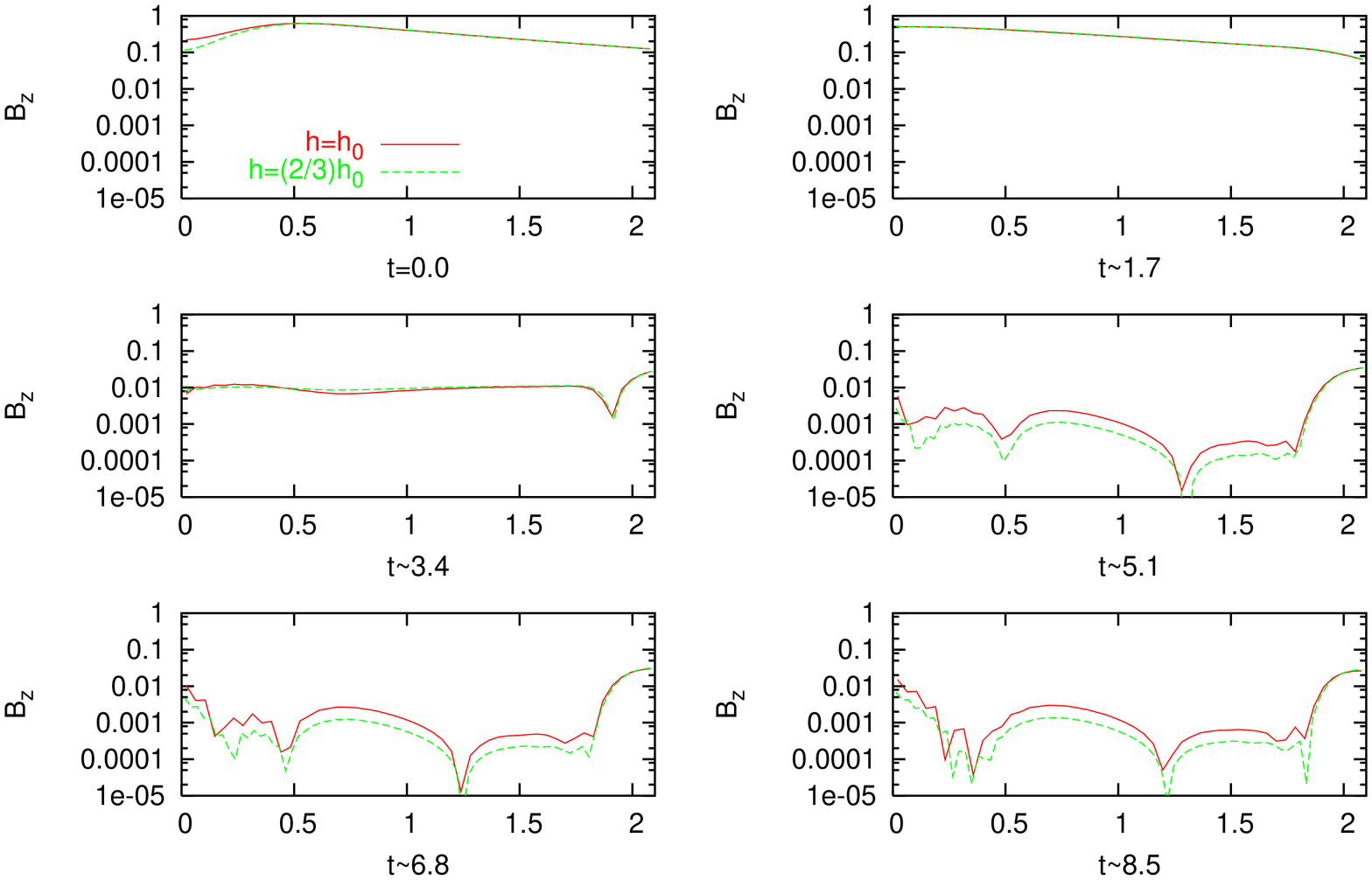,height=6in}}
\caption{A field component plotted along a line
$z=0, x=1$ parallel to the $y$-axis. At the grid limit $y=2.1$ one has
$r=2.33$, while $\scri^+$ is met at $y=1.73$. Inside $\scri^+$ the field
values appear to be convergent to zero after $u=4$, while in the region
beyond $y=1.85$ or $r=2.1$ a slowly moving pulse appears to remain
independent of the selected resolution.}
\label{fig:S-Bz_x1z0}
\end{figure*}

With a spherical excision boundary and causal propagation of all
modes, there is less room for bad initial conditions to intrude, 
and better reason to think that even poor initial conditions 
would be quickly flushed out of the computational domain.
We have simply used as initial conditions the formulas for $\mathcal{B}$
and $\mathcal{D}$ as computed from equations~(\ref{eq:A-example}) through
(\ref{eq:f_init-example}) extended analytically by the coordinate
transformation from $TSYZ$ to $uxyz$ of equations~(\ref{eq:AnMR}).
As a solution of the Maxwell equations in the Minkowski metric, these
formulae give substantial activity in the $r>2$ region, as seen in
Fig.\ref{fig:S-exact_log_incoming40}.
In that Figure one solution has been modified by smoothly deleting the 
terms involving $\exp{-\lambda V^2}$ for $r\geq 2$ which is possible 
since $V/s=u-2+2/(1-\half r)$ becomes infinite at $r=2$. 
However the pulse seen moving inward from beyond the grid will have
depended upon initial conditions far outside $r=3.64 =2.1\sqrt{3}$ which
was the limit of our grid. 
Our initial data, however, are not evolved using the Minkowski metric,
but usually with the de Sitter or modified de Sitter metrics, where the
decomposition into ingoing and outgoing parts of the initial conditions
will differ seriously from the Minkowski case, especially well beyond
$\scri^+$.
Thus we can understand that different, but equally visible, activity 
could occur in the $r>2$ region which would be actually solving the 
given differential equations but be unrelated to the physical activity 
in the region $r\le 2$.
This appears to be occuring in the numerical results (modified de Sitter 
for which no analytic solution is available) shown in Fig.~\ref{fig:S-Bz_x1z0}.



\end{document}